\documentclass[english,aps,preprint,prd,showpacs,groupedaddress,graphics]{revtex4}
\usepackage{amsmath}
\usepackage{graphicx}
\usepackage{babel}

\begin{document}

\title{Ultrarelativistic Bose-Einstein Gas on Lorentz Symmetry Violation}
\author{J.A. de Sales, T. Costa-Soares and V.J. Vasquez Otoya }
\affiliation{Instituto Federal de Educa\c{c}\~ao, Ci\^encia e Tecnologia do Sudeste de Minas Gerais - IF Sudeste MG
\\
Campus Juiz de Fora - N\'ucleo de F\'isica\\
36080-001 Juiz de Fora, Minas Gerais, Brazil }
\date{\today}

\begin{abstract}

In this paper we study the effects of Lorentz Symmetry Breaking on thermodynamics properties of ideal gases. Inspired in the dispersion relation came from the Carroll-Field-Jackiw model for Electrodynamics with Lorentz and CPT violation term, we compute the  thermodynamics quantities for a Boltzmann, Fermi-Dirac and Bose-Einstein distributions. Two regimes are analyzed: the non-relativistic and the relativistic one. In the first case we show that the topological mass induced by the Chern-Simons term behaves as a chemical potential. For the Bose-Einstein gases it could be found a condensation in both regimes, being the appearance  of a condensate in the relativistic regime, the main contribution of this work. 

\end{abstract}

  
\maketitle

\section*{INTRODUCTION}

\bigskip Since the advent of the String Theories as a candidate to the
Unified Theory, Lorentz- and CPT- Violations are expected at Planck Scale
\cite{Samuel}, and a background anisotropy on the space-time must correct
the physics at a low energy scale. For instance, at a Standard Model scale,
the Extended Standard Model SME \cite{Colladay} was proposed as a possible
extension of the minimal Standard Model of the fundamental interactions.
Even though the expectation of the Lorentz- and CPT violation, these effects
are very small, and the SME has also been used as a framework to get
stringent bounds on the Lorentz-symmetry violating (LV) coefficients \cite
{Tests}, \cite{CPT}. In this framework, there are a large number of results
in the literature that investigate these effects in  different situations,
like system involving photons\cite{photons1}, \cite{photons2}, radiative
corrections \cite{Radiative}, fermions \cite{fermions}, neutrinos \cite
{neutrinos}, topological defects \cite{Defects}, topological phases \cite
{Phases}, cosmic rays \cite{CosmicRay}, supersymmetry \cite{Susy}, particle
decays \cite{Iltan}, and other relevant aspects \cite{Lehnert1}, \cite
{General}.

This violation can be implemented in the fermion sector, for example, by $
v_{\mu }\overline{\psi }\gamma ^{\mu }\psi ,b_{\mu }\overline{\psi }\gamma
_{5}\gamma ^{\mu }\psi $, leading to a modified Dirac theory \cite
{Hamilton}. It also has consequences in \ a very low energy regime, like in atomic
physics, condensed matter and so on, taking into account the non
relativistic limit, in order to obtain experimental bounds on the Lorentz symmetry braking (LSB) parameters and
other effects, as the generation of an anomalous magnetic moment by a
non-minimal coupling covariant derivative \cite{Manojr}, \cite{Nonmini}.

In the gauge sector, it has been introduced by the pioneering work by
Carrol, Field and Jackiw, modifying the Electromagnetic Maxwell Lagrangian
by means of a Chern-Simons type extra term $(\propto \epsilon ^{\mu
\alpha \beta \gamma }v_{\mu }F_{\alpha \beta }A_{\gamma })$. This modify the
dispersion relation of the photon introducing a topological mass and
preserving the gauge invariance. It has been pointed out by Perez-Victoria
\cite{Radiative} that this term is a one loop quantum correction of the
fermion coupling to the background.

Even though LSB is a theoretical implication, the search for a deviation on
the photon dispersion relation has important consequences on cosmic physics,
like Gamma Ray Bursts, near stars with strong magnetic field and vacuum
birefringence. This can justify the study of the statistical
behavior of the photon gas under a non-linear dispersion relation. As this
violation is expected to be observed only at ultra high energy, this address
to the question of its statistical behavior in a ultra-relativistic photon
gas, not only to get more stringent bounds on the LSB parameters but also an
interesting phenomenon that could be related by this breaking.

Another interesting point is related to the light bosons \cite{light-bosons}, which are described by a scalar
field, when minimally coupled to gravity it could be a candidate for Dark
Matter. Its mass is constrained to the order $10^{-22}eV$. In ref. \cite{light-bosons}, the authors argue that the neutrino radiation
behavior is linked to a ultra-relativistic transition. This scalar field falls
in the classification of Hot Dark Matter (HDM), in the sense that it behaves as
radiation at its decoupling epoch. This  is related to a Bose-Einstein
condensation in ultra-relativistic gas. These facts motivate the study of a 
Bose-Einstein gas with a non-linear dispersion relation at the ultra-relativistic level.

In this paper, we study the statistical mechanics of an ideal gas embedded
in a LSB background, starting from a massive
Chern-Simons dispersion relation. Two regimes are computed. In the non relativistic, when the background is larger than the momentum of the
particle, we will show that in this limit our results are the same
obtained when started by the SME Lagrangian where a non-relativistic
Hamiltonian is computed for the free the Bose-Einstein, Fermi Dirac and
Boltzmann gas \cite{Colladay:2004qt}. Our contribution in this \ paper is the study of
the relativistic gas. We shall see, for the Bose gas, that LSB background can induce a phase transition for the high
energy gas.

\section{Boltzmann's gas}

As a first step, we analyze the Statistical Mechanics for an ideal Boltzmann gas, by computing the relativistic and the
non-relativistic limit. The gas is described by the dispersion relation
raised from Carroll-Field-Jackiw model, which is given by, $
p^{4}+v^{2}p^{2}-(v.p)^{2}=0$, where $\nu $ is a four-vector background
field. This background vector violates CPT and Lorentz symmetries in the
particle frame. We take the space-like vector that obey the causal
conditions described in \cite{BaetaScarpelli:2003kx}. In this regime, $E$ is
given by

\begin{equation}
E=\sqrt{\frac{2\vec{p}^{2}+\vec{v}^{2}}{2}\pm \frac{1}{2}\sqrt{\vec{v}^{4}+4(
\vec{v}.\vec{p})^{2}}}.  \label{disp-relat}
\end{equation}

In statistical mechanics the partition function for a system of N particles
is given by $\ $
\begin{equation}
\Xi \left( T,V,\mu \right) =\sum_{N=0}^{\infty }z^{N}Z\left( T,V,N\right)
\label{Boltz-Parti}
\end{equation}
where $Z\left( T,V,N\right) =\frac{1}{N!}Z^{N}\left( T,V,1\right) $ and $
Z\left( T,V,1\right) =\frac{1}{h^{3}}\int e^{-\beta E(p,q)}dpdq$ with $\beta
=\frac{1}{kT}$ and $z=e^{\beta \mu }$. In order to split the two regimes
that we are interested in, we analyze the limit cases, when $|\vec{v}|\gg |
\vec{p}|,$ that is the non-relativistic regime of the (\ref{disp-relat})

\begin{equation}
E_{\pm }=|\vec{v}|+\frac{1}{2}\frac{p^{2}}{|\vec{v}|}(1\pm \cos ^{2}\theta )
\label{non-relat}
\end{equation}
and the relativistic case, $|\vec{v}|\ll |\vec{p}|,$ the first order on $v$
of (\ref{disp-relat}) assumes
\begin{equation}
E=|\vec{p}|+\frac{1}{2}|\cos \theta |v,  \label{relat}
\end{equation}

\subsection{The non-relativistic Limit: $|\vec{v}|\gg |\vec{p}|$}

In order to study the statistical mechanics of the classical gas in the
non-relativistic limit, we put the dispersion relation (\ref{non-relat}) in
the partition function (\ref{Boltz-Parti}). Here, we will not consider the
case with negative energy, by means that this situation violates the
causality conditions, \cite{BaetaScarpelli:2003kx}. Then, 

\begin{equation}
Z\left(T,V,1\right) =\frac{2\pi Ve^{-\beta v}}{h^{3}}\int_{0}^{\pi
}\int_{0}^{\infty }\sin \theta e^{-\beta \lbrack \frac{1}{2}\frac{p^{2}}{v}
(1+cos^{2}\theta )]}p^{2}dpd\theta, 
\end{equation}
after a straightforward integration,
we obtain the $Z-$function $Z\left( T,V,1\right) =\frac{1}{4}\left( \frac{2v
}{2\pi \beta }\right) ^{3/2}Ve^{-\beta v}$.

Thus, the partition function is

\begin{equation}
\Xi\left(T,V,\mu\right)=\sum_{N=0}^{\infty}\frac{1}{N!4^{N}}\left(\frac{2v}{
2\pi\beta}\right)^{\frac{3}{2}N}V^{N}e^{\beta\left(\mu-v\right)N}.
\end{equation}

To compute the number of particles $N=-\frac{\partial \phi }{
\partial \mu }$, pressure $P=-\frac{\partial \phi }{\partial V}$ and the
entropy $S=-\frac{\partial \phi }{\partial T}$, we calculate the
grand-potential function
\begin{equation}
\phi =-kT\ln \Xi =-kTe^{\beta \left( \mu -v\right) }\frac{1}{4}\left( \frac{
2v}{2\pi \beta }\right) ^{3/2}V,
\end{equation}
that results, respectively 

\begin{eqnarray}
N &=&e^{\beta \left( \mu -v\right) }\frac{1}{4}\left( \frac{2v}{2\pi \beta }
\right) ^{3/2}V,  \notag \\
P &=&\frac{kTN}{V},  \notag \\
S &=&kN\left[ \frac{5}{2}-\left( \mu -v \right) \beta \right] .
\end{eqnarray}


The energy can be computed by the relation, $U=TS-PV+\mu N$ which results in
$U=\frac{3}{2}NkT+vN$, and the state equation $U=\left( \frac{3}{2}+\beta
v\right) PV$. 
The chemical potential can be calculated explicitly in this case as

\begin{equation}
\mu=-kT \left[\ln \left( \frac{N_Q}{N_C}\right)-(\frac{v}{kT}+\ln 4 -1)\right],
\end{equation}
where we call the quantities $N_Q =\left(\frac{2v}{2\pi\beta}\right)^{\frac{3}{2}}$ as the quantum concentration and $N_C=\frac{N}{V}$ the 
standard classical concentration, or density of particles. Note that the chemical potential is affected by the background field $v$ as was 
reported by Colladay \textit{et al} \cite{Colladay:2004qt}. This change is due to the behavior of the field as an effective mass for the gas.

In the thermodynamic limit, and  when $kT\gg v$, we recover the
non-relativistic limit result to the photon gas without Lorentz-violation
and the specific heat 

\begin{equation}
C_{V}=\frac{\partial U}{\partial T}=\frac{3}{2}Nk.
\end{equation}%
Thus, our approach recover the standard results when the background vector $v$ is neglected for the non relativistic
gas. In this situation, it is not interesting to investigate bounds on the
LSB parameters because we are describing here a classical non-relativistic
photon gas.

\subsection{The Relativistic Limit $|\vec{v}|<<|\vec{p}|$,}

By the same procedure used in the non-relativistic limit, we find for the $Z$
-function $Z\left( T,V,1\right) =\frac{8V}{\left( 2\pi \right) ^{2}\beta
^{4}v}e^{-\frac{\beta v}{2}}(e^{\frac{\beta v}{2}}-1)$, that the partition
function and the grand potential become%
\begin{eqnarray}
\Xi \left( T,V,\mu \right) &=&\exp \left[ e^{\beta \mu }\frac{8V}{\left(
2\pi \right) ^{2}\beta ^{4}v}e^{-\frac{\beta v}{2}}(e^{\frac{\beta v}{2}}-1)
\right]  \notag \\
\phi &=&-kT\left[ e^{\beta \mu }\frac{8V}{\left( 2\pi \right) ^{2}\beta ^{4}v
}e^{-\frac{\beta v}{2}}(e^{\frac{\beta v}{2}}-1)\right]
\end{eqnarray}
Then the  particle number is 

\begin{eqnarray*}
N&=&e^{\beta \mu }\frac{8V}{\left( 2\pi \right) ^{2}\beta ^{4}v}e^{-\frac{
\beta v}{2}}(e^{\frac{\beta v}{2}}-1).
\end{eqnarray*}

The entropy, energy and the specific heat $C_{V}$ with the background are
respectively
\begin{eqnarray}
S &=&kN\left[ 5-\mu \beta -\frac{v}{2}\beta -\frac{v}{2}\beta \left( \frac{
e^{\frac{\beta v}{2}}}{e^{\frac{\beta v}{2}}-1}\right) \right]
\label{Entropy-Boltz} \\
U &=&\left[ 4-\left( \frac{\frac{\beta \nu }{2}}{e^{\frac{\beta \nu }{2}}-1}
\right) \right] PV  \label{State-Boltz} \\
C_{V} &=&4kN-\frac{e^{\frac{v}{2kT}}Nv^{2}}{4(e^{\frac{v}{2kT}}-1)kT^{2}}.
\label{CV-Boltz}
\end{eqnarray}
and we have the same state equation $PV=NkT$.  

The chemical potential is now given by 

\begin{equation}
\mu=kT \left[ \ln\left(\frac{\pi^2\beta^4 v}{2 N_C}\right)-\left(\frac{v}{kT}+\ln \left(1-e^{-\frac{\beta v}{2}}\right)-1 \right)\right].
\end{equation}
We can observe how the background parameter $v$ modify the statistics.
In the limit which the background does not exist, we recover the standard
result $ U=3NkT $ and $C_v=3Nk$. The behavior of the Boltzmann gas in the presence of background field is illustrated in the figures 
that follow. In Fig.1 we can observe the behavior of the internal energy with the temperature and a fixed parameter $v$. For large temperatures all the curves tend asymptotically to the value $U=3NkT$ and collapse to this value when the parameter v goes to zero. The same behavior can be seen in Fig.2, where $U$ is a function of $v$. Note here that the field $v$ clearly introduces an effective mass in the system so that the internal energy increases. The behavior of the specific heat is shown in Fig.3 and again we found the change introduced by the background field. When $v$ goes to zero at $T = 0$ the value of $C_v=3Nk$, the same expected by Boltzmann's  statistics.


\section{Fermi-Dirac Statistics}

We now begin to study the effects of the LSB background on the quantum
gases. To this end, we should compute the Fermi-Dirac and Bose-Einstein
distribution and in this paper we are interested in both limits,
non-relativistic and relativistic regime. In the Fermi-Dirac statistics the
grand potential is given by

\begin{equation}
\phi =-kT\,\frac{2\pi V}{h^{3}}\frac{\beta }{3}\int_{0}^{\pi }\int_{0}^{\infty
}\sin \theta \,p^{3}\frac{ze^{-\beta \epsilon }}{1+ze^{-\beta \epsilon }}
\left( \frac{d\epsilon }{dp}\right) dp\,d\theta .
\end{equation}
The non relativistic regime, i.e. $|\vec{v}|\gg |\vec{p}|$ , $\epsilon =|
\vec{v}|+\frac{1}{2}\frac{p^{2}}{v}(1+\cos ^{2}\theta )$ and $\frac{
d\epsilon }{dp}=\frac{p}{v}(1+\cos ^{2}\theta )$, which yields the grand
partition function


\begin{equation}
\phi =-kT\,\frac{V}{\lambda ^{3}}\frac{\sqrt{2}}{2}f_{5/2}(ze^{-\beta v})
\end{equation}%

We should note that the thermal wave length of the Fermi gas  $\lambda $, depends on the LSB
parameter

\begin{equation}
\lambda =\left( \frac{h^{2}}{2\pi vkT}\right) ^{1/2}
\end{equation}
and $f_{n}(\chi )$ is the complete Fermi-Dirac integral defined by

\begin{equation}
f_{n}(\chi )=\frac{1}{\Gamma (n)}\intop_{0}^{\infty }\frac{\xi ^{n-1}}{\chi
^{-1}e^{\xi }+1}d\xi .
\end{equation}

Some important thermodynamic quantities as  particle number, pressure and
density of energy is defined below:

\begin{eqnarray}
N &=&\frac{V}{\lambda ^{3}}\frac{\sqrt{2}}{2}f_{3/2}(ze^{-\beta
v})  \notag \\
P &=&\frac{1}{\lambda ^{3}}(kT)\frac{\sqrt{2}}{2}f_{5/2}(ze^{-\beta v})
\notag \\
U &=&-\frac{1}{V}\frac{\partial \phi }{\partial \beta }=\frac{1}{
\lambda ^{3}}\frac{\sqrt{2}}{2}\left[ vf_{3/2}(ze^{-\beta v})+\frac{3}{2}
(kT)f_{5/2}(ze^{-\beta v})\right]V
\end{eqnarray}
The the equation of state for a Fermi gas is

\begin{equation*}
PV=NkT\frac{f_{5/2}(ze^{-\beta v})}{f_{3/2}(ze^{-\beta v})}.
\end{equation*}

For high temperatures the Fermi's gas behaves like a Boltzmann' s gas, then
the complete Fermi-Dirac function is $f_{n}(\chi )\approx \chi $, or $PV = NkT$. These results are the same obtained by Colladay \cite{Colladay:2004qt} 

\bigskip The relativistic regime, i.e. $|\vec{v}|\ll |\vec{p}|$ and the
energy $\epsilon =|\vec{p}|+\frac{1}{2}|\cos \theta |v$, $\ \frac{d\epsilon
}{dp}=1.$ The grand partition function can be written as



\begin{eqnarray}
\phi & = & -kT\,\frac{8\pi V}{h^{3}}\frac{1}{\beta^{3}}I_{5}(z,\chi)
\end{eqnarray}

where the functions $I_{n}$(z$\chi)$ are

\begin{eqnarray}
I_{n}(z,\chi) & = & \frac{1}{\ln\chi}\left(f_{n}(z)-f_{n}(z\chi)\right)
\end{eqnarray}

In the limit when $v\rightarrow0$ the grand potential becomes

\begin{eqnarray}
\phi & = & -kT\,\frac{8\pi V}{h^{3}}\frac{1}{\beta^{3}}f_{4}(z)
\end{eqnarray}

The pressure may be written like

\begin{eqnarray}
P & = & \frac{8\pi}{h^{3}}(kT)^{4}I_{5}(z,\chi)
\end{eqnarray}

For $v\rightarrow0$ and $kT$ $\gg\mu$, we have

\begin{equation*}
P=\frac{8\pi }{h^{3}}(kT)^{4}\frac{\pi ^{4}}{90}\frac{7}{8}
\end{equation*}

We  see that the pressure for Fermi-Dirac gas differs by the factor $%
\frac{7}{8}$ when compared to Bose gas in same condition $v\rightarrow 0$
and $z\rightarrow 1$ as usual.

The internal energy $U$ is

\begin{equation*}
U=\frac{8\pi V}{h^{3}}(kT)^{4}(4I_{5}(z,\chi )-f_{4}(z\chi ))
\end{equation*}

when $v\rightarrow 0$ and $z\rightarrow 1$
\begin{equation}
U=\frac{8\pi V}{h^{3}}(kT)^{4}\,\frac{7\pi^4}{240}
\end{equation}

\section{Bose-Einstein Statistics}

Colladay \textit{et. al }have been studied the effects of the LSB background
in a general non-relativistic statistics. The general effects of the
background in this kind of situation is to redefine the thermodynamics
quantities, unlike to obtain a new effect, since the background is expected
to be small in low energy scale. In this framework, these systems can be
used to get experimental bounds on the LSB parameters. \ In a Bose gas
Colladay \textit{et. al }have been analyzed in two very different situation
\textit{.} In a first publication, it has been shown how the background can
modify the standard thermodynamics results as the critical temperature for
the homogeneous non relativistic Bose-Einstein Condensates (BEC) phase
transition, starting from the non-relativistic Hamiltonian with LSB
non-relativistic contribution. In a second publication it has been proposed
to use the BEC trapped as probe to LSB parameters.

Here, we starts from the Carroll-Field Jackiw dispersion relation (\ref
{disp-relat}), and compute both limits. The non-relativistic regime by the
same way that was computed to the Boltzmann Gas (\ref{non-relat}). We will
show that our results are the same obtained by Colladay \textit{et. al},
using other method. Our goal here, is the study of the ultra-relativistic
free Bose gas under the same background. We will see that under this regime,
a phase transition can be induced by the vector background.

The grand canonical partition function can be calculated evaluating the
number of states in the one-particle phase space $\Sigma $, where in
spherical coordinates is $\Sigma =\frac{2\pi V}{h^{3}}\int_{0}^{\pi }\sin
\theta d\theta \int_{0}^{\infty }p^{2}dp,$the one-particle density of states
is given by%
\begin{equation}
g(\epsilon )=\frac{d\Sigma }{d\epsilon }=\left[ \frac{2\pi V}{h^{3}}
\int_{0}^{\pi }\sin \theta \int_{0}^{\infty }p^{2}dp\right] \frac{dp}{
d\epsilon }.
\end{equation}
The macro canonical partition function  $\Xi$ in Bose statistics is given by

\begin{equation}
\ln\Xi(T,V,z)=-\sum_{k}^{\infty}\ln(1-ze^{-\beta\epsilon_{k}}).
\end{equation}

By computing the thermodynamic limit, $\ln \Xi =-\int_{0}^{\infty }g(\epsilon )\ln
(1-ze^{-\beta \epsilon })d\epsilon -\ln (1-z),$ then grand potential is

\begin{equation}
\phi =-kT \,\frac{2\pi V}{h^{3}}\frac{\beta }{3}\int_{0}^{\pi }\int_{0}^{\infty
}\sin \theta \,p^{3}\frac{ze^{-\beta \epsilon }}{1-ze^{-\beta \epsilon }}
\left( \frac{d\epsilon }{dp}\right) dp\,d\theta -\ln (1-z)
\label{Bose-Grand-Potential}
\end{equation}
where $z=e^{\beta \mu }$ is the fugacity. 

\subsection{The non-relativistic Limit $|\vec{v}|\gg |\vec{p}|$:}

The dispersion relation in the non-relativistic limit is given by (\ref
{non-relat}). To study the effects of LSB in a non-relativistic Bose gas, we
put this dispersion relation in (\ref{Bose-Grand-Potential}). In this
situation, the grand potential function can be written as

\begin{equation}
\phi=-kT\,\frac{2\pi V}{h^{3}}\frac{\beta}{3}\int_{0}^{\pi}\int_{0}^{\infty}\sin
\theta\, p^{3}\frac{ze^{-\beta\left[v+\frac{1}{2}\frac{p^{2}}{v}
(1+\cos^{2}\theta)\right]}}{1-ze^{-\beta\left[v+\frac{1}{2}\frac{p^{2}}{v}
(1+\cos^{2}\theta)\right]}}\frac{p}{v}(1+\cos^{2}\theta)dp\, d\theta
\end{equation}
after evaluating the integral, we obtain 

\begin{equation}
\phi =-kT\,\frac{V}{\lambda ^{3}}\frac{\sqrt{2}}{2}g_{5/2}(e^{\beta (\mu-v)})
\end{equation}
where $g_{n}\left( z\right) ,$ is defined by $g_{n}\left( z\right) =\frac{1}{
\Gamma \left( n\right) }\int_{0}^{\infty }\frac{x^{n-1}dx}{z^{-1}e^{x}-1}
,\qquad 0\leq z\leq 1$and $\lambda =\left( \frac{h^{2}}{2\pi vkT}\right)
^{1/2}=\left( \frac{2\pi }{vkT}\right) ^{1/2},$is the thermal wavelength.
\newline
Computing the thermodynamical quantities, we obtain

\begin{eqnarray}
P &=&\frac{1}{\lambda ^{3}}(kT)\frac{\sqrt{2}}{2}g_{5/2}(e^{\beta (\mu-v)}),
\notag \\
N &=&\frac{V}{\lambda ^{3}}\frac{\sqrt{2}}{2}g_{3/2}(e^{\beta(\mu- v)})+N_{0}
\end{eqnarray}
The total number N is then $N=N_{1}+N_{0}$ where $N_{0}$ is the number of
particles in the ground state and $N_{1}=\frac{V}{\lambda ^{3}}\frac{\sqrt{2}
}{2}g_{3/2}(e^{\beta (\mu-v)})$ is the number of particles in excited states.

The critical temperature $T_{c}$ when the Bose-Einstein condensation (BEC)
occurs is obtained when $g_{3/2}(x)$ has a maximum, or $\mu=v$.

\begin{equation}
T_{c}=\left( \frac{N}{V}\right) ^{2/3}\frac{h^{2}}{2\pi vk}\frac{1}{(2\zeta
(3/2)^{2})^{1/3}},
\end{equation}
the number of particles bellow of the critical temperature, or $T<T_{c}$ is

\begin{equation}
N=N_{1}=\frac{V}{\lambda ^{3}}\frac{\sqrt{2}}{2}\zeta (3/2).
\end{equation}

For $T>T_{c}$ we have

\begin{equation}
N=N_{1}=\frac{V}{\lambda ^{3}}\frac{\sqrt{2}}{2}g_{3/2}(e^{\beta(\mu- v})),
\end{equation}
the remaing particles are in ground state, so we can write

\begin{equation}
\frac{N}{N_{0}}=\frac{N-N_{1}}{N}=1-\left( \frac{T}{T_{c}}\right) ^{3/2}.
\label{eq:temp-crit}
\end{equation}

The internal energy $U$ for $T<T_{c}$ is

\begin{equation}
U=N_{1}(kT)\left[ \frac{v}{kT}+\frac{3}{2}\frac{g_{5/2}(1)}{g_{3/2}(1)}
\right] ,
\end{equation}
Using the equation eq. (\ref{eq:temp-crit}) the specific heat is $C_{V}=\frac{
15}{4}Nk\frac{\zeta (5/2)}{\zeta (3/2)}\left( \frac{T}{T_{c}}\right) ^{3/2}$. 
Above of the critical temperature, $T>T_{c}$, the internal energy $U$
assumes

\begin{equation}
U=N(kT)\left[ \frac{3}{2}\frac{g_{5/2}(z)}{g_{3/2}(z)}\right] ,
\end{equation}
where $z$ is now set as $z=e^{\beta( \mu-v)}$. The specific heat reads

\begin{equation}
C_{V}=\frac{3}{2}Nk\left[ \frac{5}{2}\frac{g_{5/2}(z)}{g_{3/2}(z)}-\frac{3}{2
}\frac{g_{3/2}(z)}{g_{1/2}(z)}\right]
\end{equation}

\subsection{The Ultra-relativistic Bose Gas: $|\vec{p}|\gg |\vec{v}|$}

In this section, the ultra-relativistic ideal Bose gas is studied,
starting from the dispersion relation Carrol-Field-Jackiw (\ref{relat}) in
the Bose-Einstein distribution \ref{Bose-Grand-Potential}. We will see that
the background modifies the thermodynamics of the gas, introducing a phase
transition that does not exist in the standard Bose gas. This effect is very
interesting and this result could be important for astrophysical and cosmological applications \cite{urenha, light-bosons}.

The grand potential is
\begin{equation}
\phi =-kT \,\,\frac{2\pi V}{h^{3}}\frac{\beta }{3}\int_{0}^{\pi }\int_{0}^{\infty
}\sin \theta \,p^{3}\frac{ze^{-\beta \left[ p+\frac{1}{2}|\cos \theta |
\right] }}{1-ze^{-\beta \left[ p+\frac{1}{2}|\cos \theta |v\right] }}
dp\,d\theta
\end{equation}
Performing the integration over $\theta $and $p$  we obtain

\begin{equation}
\phi =-kT \,\,\frac{8\pi V}{h^{3}}\frac{1}{\beta ^{3}}\left[ \frac{2}{\beta v}\left(
g_{5}(e^{-{\beta \mu}})-g_{5}(e^{-{\beta( \mu-\frac{v}{2}}})\right) \right] .
\end{equation}

In the limit $v\longrightarrow 0$ and $\mu=0$ we have the standard photon gas, the function $g_{5}(e^{-\frac{\beta v}{2}
})\longrightarrow \zeta (5)$, and the grand potential for the
ultrarelativistic Bose-Einstein gas, in the limit where the LSB background
does not exist, becomes the well-known result

\begin{equation}
\phi =-kT\,\,\frac{8\pi V}{h^{3}}\frac{1}{\beta ^{3}}g_{4}(1),
\end{equation}%
with $g_{4}(1)=\frac{\pi ^{4}}{90}$.

For further calculations we will write now the grand potential as

\begin{equation}
\phi =-kT \,\,\frac{8\pi V}{h^{3}}\frac{1}{\beta ^{3}}F_{5}(z,v ),
\end{equation}%
where the function $F_{5}(z,v )$ will be definite by 

\begin{equation}
F_{5}(z,v )=\frac{2}{\beta v}\left(g_5(z)- g_5(ze^{-\beta \frac{v }{2}})\right)
\end{equation}
and is easy to show that its derivative  with respect to $\beta $ is given by 

\begin{equation}
\dfrac{\partial F_{5}(z,v)}{\partial \beta}= \frac{1}{\beta }\left(g_4(ze^{-\beta\frac{v}{2}})- F_5(z,v)\right).
\end{equation}

The thermodynamical quantities are given by

\begin{equation}
P=\frac{8\pi }{h^{3}}(kT)^{4}F_{5}(z,v ).
\end{equation}
When $v\longrightarrow 0$ we obtain the well-known result for ultrarelativistic
Bose-Einstein gas $P=\frac{8\pi }{h^{3}}(kT)^{4}\frac{{\pi }^{4}}{90}$.

The internal energy can be calculated by

\begin{equation}
U=\frac{8\pi V}{h^{3}}(KT)^{4}\left[ 4F_{5}(z,v )-g_4(ze^{-\beta\frac{v}{2}})\right] ,
\end{equation}
the state equation is then

\begin{equation}
\frac{U}{V}=\frac{4F_{5}(z,v) - g_4(ze^{-\beta\frac{v}{2}}) }{F_{5}(z,v) }P.
\end{equation}

When $v\longrightarrow 0$ the energy density is

\begin{equation}
\frac{U}{V}=3P
\end{equation}

In the above equations we can observe that the pressure and the energy
density depends only of the temperature and it vanishes as $T^{4}$ for $%
T\longrightarrow0$. In PV diagrams the isotherms are parallel lines to V-axis.




\subsection{The Mean Particle Number and the Bose Temperature}

A very useful quantity is the mean particle number, that in Bose-Einstein
statistics is given by $N(T,V,z)=\sum_{k}\frac{1}{z^{-1}e^{\beta \epsilon
_{k}}-1}$. In the thermodynamic limit we have

\begin{equation}
N=\frac{2\pi V}{h^{3}}\int_{0}^{\pi}\int_{0}^{\infty}\sin\theta\, p^{2}\frac{
ze^{-\beta\left[p+\frac{1}{2}|\cos\theta|v\right]}}{1-ze^{-\beta\left[p+
\frac{1}{2}|\cos\theta|v\right]}}dp\, d\theta
\end{equation}

or explicitly 

\begin{equation}
N=\frac{8\pi V}{h^3}\frac{1}{\beta^3}\frac{2}{\beta v}\left( g_4(z)-g_4(ze^{-\beta \frac{v}{2}})\right)
\end{equation}

 The maximum mean number of particle is defined when $ \mu = \frac{v}{2}$ or when we reach the critical temperature $T_c$. In that temperature we have 
 
\begin{equation}
N=\frac{8\pi V}{h^3}\frac{1}{\beta^3}\frac{2}{\beta v}\left( g_4(e^{-\beta \frac{v}{2}})-\zeta (4)\right).
\end{equation}

Expanding the above equation for $\frac{v}{2kT} \ll 1$  and  taking only terms in first order of $v$ we have

\begin{equation}
N=\frac{8\pi V}{h^{3}}(kT)^{3}\left(\zeta(3)+\frac{\pi^2}{12}\,\frac{v}{2kT}\right)
\label{t-crit}
\end{equation}

The critical temperature $Tc$, where the Bose Einstein 
Condensation occurs is the real cube root of the Eq.(\ref{t-crit}). The necessary condition for  $Tc$  real  is 

\begin{equation}
v\le \frac{18 (16\pi^2\zeta(3)^2)^{1/3}}{\pi^2•}\frac{\hbar}{c}\, \left(\frac{N}{V}\right)^{1/3},
\end{equation}
where $N/V$ is the density of the condensate. The maximum mass of $v$ so that the  gas is a condensate in $MeV/c^2$ is 

\begin{equation}
v\le 0.049465\, \rho^{1/4},
\end{equation}
where $\rho$ is now the density of the gas in $kg/m^3$.

When $T=Tc$ the energy of the gas 

\begin{equation}
U=NkT\left( \frac{4F_{5}(z,v)-g_4(ze^{-\beta \frac{v}{2}})}{F_{4}(z,v )}\right)
\end{equation}

becomes

\begin{equation}
U=N_1kT\left\{ \frac{4(g_5(e^{\beta\frac{v}{2}})-\zeta (5))-\frac{\beta v}{2}\zeta(4)}{g_4(e^{\beta \frac{v}{2}})-\zeta(4)}\right\}
\end{equation}

and the specific heat is given by

\begin{equation}
C_v=4Nk\left(\frac{T}{T_c}\right)^3\left\{ \frac{4(g_5(e^{\beta\frac{v}{2}})-\zeta (5))-\frac{\beta v}{2}\zeta(4)}{g_4(e^{\beta \frac{v}{2}})-\zeta(4)}\right\}+ (kT)\left(\frac{T}{T_c}\right)^3\frac{\partial}{\partial T}\left\{ \frac{4(g_5(e^{\beta\frac{v}{2}})-\zeta (5))-\frac{\beta v}{2}\zeta(4)}{g_4(e^{\beta \frac{v}{2}})-\zeta(4)}\right\}
\end{equation}

 The standard case (photon gas and absence of background field) is described in the limit $v\rightarrow0 $.

\begin{equation}
\frac{C_{V}}{Nk}=\frac{3\zeta (4)}{\zeta (3)}\approx 2.70118
\end{equation}

\subsection{The BEC of an Ultrarelativistic Bose Gas with a Conserved Quantum Number}

The behavior of the Bose-Einstein condensate in high temperatures when we have an ideal gas with a conserved quantum number, or generically referred as  ``charge'' is quite different from that studied so far. Haber and Weldon \cite{howard} for the first time  showed  the effects  on  the critical temperature when antiparticles are introduced in the theory.

The net charge $Q$ of the Bose gas is given by the expression 

\begin{equation}
Q=V \sum_k \left[ \frac{1}{e^{\beta(E_k -\mu)}} - \frac{1}{e^{\beta(E_k + \mu)}}•\right]
\label{eq-charge}
\end{equation}

In our case, using the dispersion relation given in  Eq. (\ref{relat}) and taking the thermodynamic limit in Eq. (\ref{eq-charge}), we find

\begin{equation}
\rho=\frac{N}{V}=\frac{2\pi}{h^{3}}\int_{0}^{\pi}\int_{0}^{\infty}\sin\theta \, p^{2}\left(\frac{1}{e^{\beta\left[p+\frac{1}{2}|\cos\theta|v-\mu \right]}- 1} - \frac{1}{e^{\beta\left[p+\frac{1}{2}|\cos\theta|v+\mu \right]}- 1} \right) dp\, d\theta
\end{equation}

After integrating over p and $\theta$ we obtain the explicit result in terms of poly-logarithm functions

\begin{equation}
\rho= \frac{8\pi}{h^3} (kT)^3 \frac{2}{\beta v}\left[g_4(e^{\beta \mu})- g_4(e^{-\beta \mu}) -g_4(e^{\beta(\mu-v/2)})+g_4(e^{-\beta(\mu+v/2)})\right].
\end{equation}

In the regime of high temperatures when the pair creation is very favorable, $ \frac{v}{2kT•}\ll 1$. The BEC occurs when $\mu=\frac{v}{•2}$, and net density of charge for the critical temperature $T_c$ reads

\begin{equation}
T_c=\sqrt{\frac{h^3}{4\pi^3k^2}\frac{3|\rho|}{v}}\label{tcritica}
\end{equation}

This result is the same obtained for the authors \cite{howard} in their seminal work, two decades ago. The new critical temperature depends  now directly  of the parameter $v$.

\section{Final Comments}

In this paper we have analyzed how the Lorentz Symmetry breaking background
modify the statistical behavior of a many particle system starting from the
Carrol-Field-Jackiw dispersion relation. We have computed particles in
classical and quantum regime, and special attention was devoted to
Bose-Einstein statistics in relativistic and non-relativistic approach. It
was pointed out that in the non-relativistic regime our approach gives raise
the same results obtained by Colladay \textit{et al}, where the
thermodynamics quantities must be corrected by the presence of $v^{\mu }$.
These results open up the possibilities in bounds on the LSB parameters.

An interesting scenario is the ultra-relativistic thermodynamic approach,
where $v^{\mu }$ induce a relativistic phase transition of a system with
bosons. In the paper of ref \cite%
{light-bosons}, it is argued that the radiation behavior of hot dark matter
is close related to a relativistic Bose-Einstein phase transition of a
scalar charged field when coupled to gravity, even though this relation is
not very well understood. \cite{urenha}. The similarities in the behavior of
SFDM BEC and the Ultra-Relativistic Bose gas under a LSB background could be
pointed out as the LSB parameter that is responsible to the phase transition
is also very small. In our case, the origin of this term is completely
different as well as it is not scalar field, but vectorial. It should be pointed out that the mass of these particles are related to the medium, in the same sense of in condensed matter. A background renormalizes the photon mass, ans it works like an effective mass. 

It is interesting to note that the critical value of the energy density of condensate is related to the LSB parameter by $v\le 0.049465\, \rho^{1/4}$. This could be used for the different universe radiation eras to fix the LSB parameter taking account the red-shift of the quantities due to the big-bang expansion.  

Another important result is about the Eq.(\ref{tcritica}), which relates the critical temperature with the LSB parameter. This could be used to calculate the density of energy and pressure of different species of particles.

\section{Acknowledgments}

\noindent J. A. Helayel-Neto and R. F. Sobreiro  are kindly acknowledged for long
discussion.

\newpage
\begin{figure}[!htb]
\includegraphics[scale=0.5]{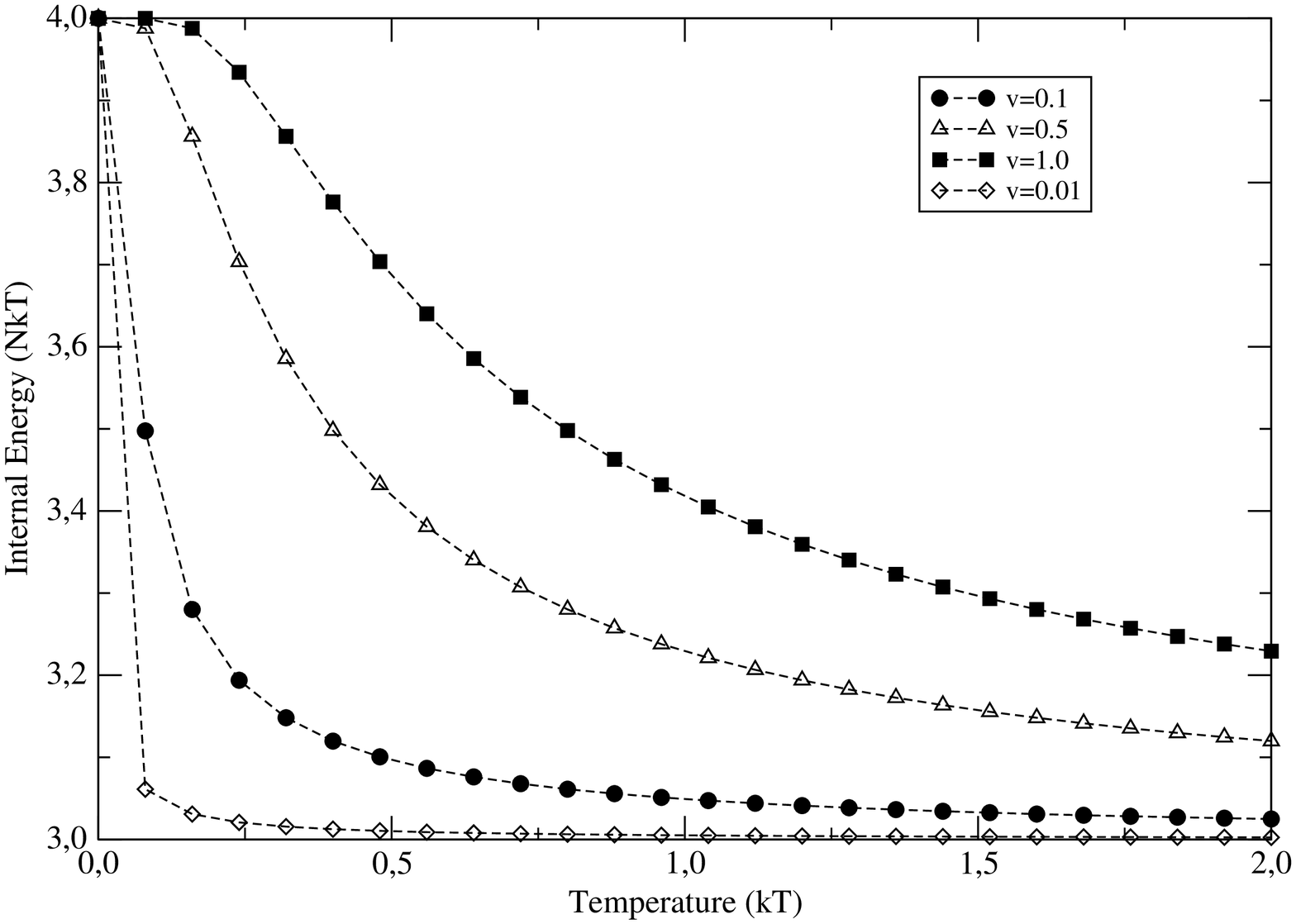}
\caption{ Internal Energy $U$ as a function of kT. We used values v = 1.0,
0.5, 0.1, 0.01. When $v \rightarrow 0$ all curves will degenerate to $U= 3
NkT$ for any temperature. }
\label{autonum}
\end{figure}


\begin{figure}[!htb]
\includegraphics[scale=0.5]{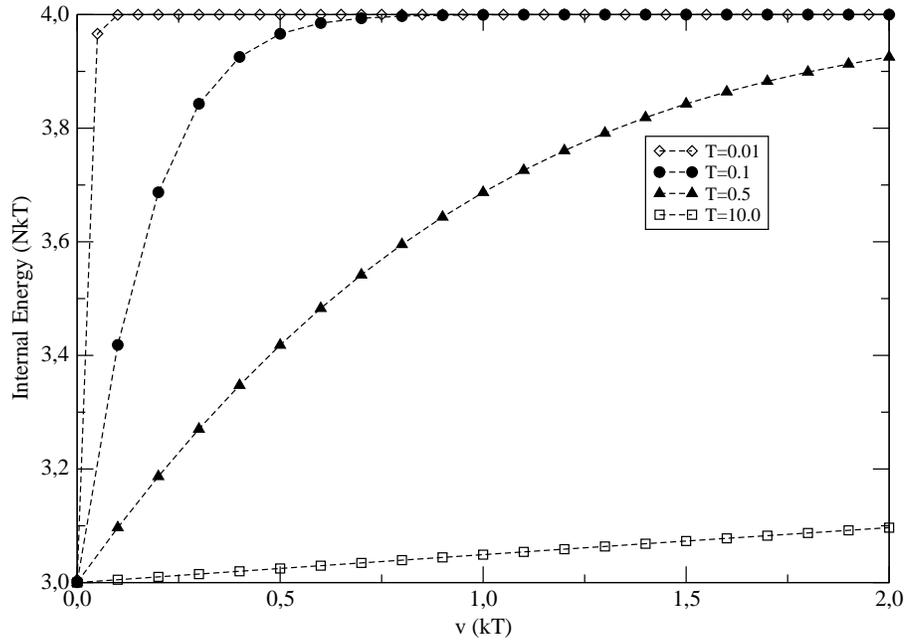}
\caption{ Internal Energy $U$ as a function of v. We used values T = 0.01,
0.1, 0.5, 10.0 . When $v\rightarrow0$ all curves converge to $U=3NkT$ for
any temperature.}
\label{autonum}
\end{figure}

\begin{figure}[!htb]
\includegraphics[scale=0.5]{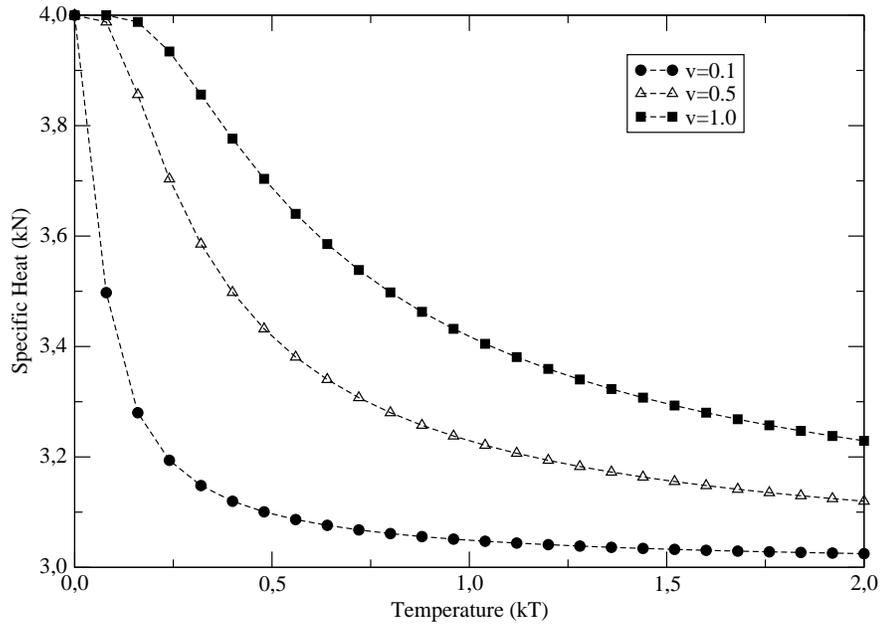}
\caption{ Specific Heat $Cv$ as a function of T. We used values v = 1, 0.5,
0.1 .When $v\rightarrow0$ all curves converge to $C_{v}=3kN$ for any
temperature. }
\label{autonum}
\end{figure}

\end{document}